\documentclass[amsmath,amssymb,aps,eqsecnum,graphicx]{revtex4}
\usepackage[dvips]{graphicx}

\begin{document}
\title{Bures volume of the set of mixed quantum states}
\author{Hans-J{\"u}rgen Sommers$^1$ and Karol {\.Z}yczkowski$^{2,3}$}

\affiliation{$^1$Fachbereich Physik,
Universit\"{a}t Duisburg-Essen, Standort Essen, 
  45117 Essen, Germany}

\affiliation {$^2$Instytut Fizyki im. Smoluchowskiego,
Uniwersytet Jagiello{\'n}ski,
ul. Reymonta 4, 30-059 Krak{\'o}w, Poland}
\affiliation{$^3$Centrum Fizyki Teoretycznej, Polska Akademia Nauk,
Al. Lotnik{\'o}w 32/44, 02-668 Warszawa, Poland}

  \date{February 17, 2005}

\begin{abstract}
We compute the volume of the $N^2-1$ dimensional set
${\cal M}_N$ of density matrices of size
$N$ with respect to the Bures measure and show that
it is equal to that of a $N^2-1$ dimensional hyper-hemisphere
of radius $1/2$.
For $N=2$ we obtain the volume of the Uhlmann hemisphere,
$\frac{1}{2} {\bf S}^3 \subset {\mathbb R}^4$.
We find also the area of the boundary of the set  ${\cal M}_N$ 
and obtain analogous results for the smaller set of all
real  density matrices. An explicit formula for
the Bures-Hall normalization constants is derived for
an arbitrary $N$.

\end{abstract}


\maketitle

\medskip
\begin{center}
{\small e-mail: sommers@theo-phys.uni-essen.de
  \ \quad \ karol@cft.edu.pl}
\end{center}


\section{Introduction}

Modern applications of quantum mechanics renewed interest in
the properties of the set of density matrices of finite size.
Its geometry depends on the metric used \cite{PS96,ZSlo01}.
However, the simplest possible
 choices exhibit some drawbacks:
the trace metric, defined by the trace distance
\begin{equation}
D_{\rm Tr}(\rho,\sigma)=||\rho-\sigma||_1:=
{\rm Tr}|\rho-\sigma|=
{\rm Tr}\sqrt{(\rho-\sigma)^2},
\label{trace}
\end{equation}
is monotone \cite{Ru94}, but not Riemannian, while
the Hilbert--Schmidt metric,  induced by the Hilbert--Schmidt scalar
product, $\langle A |B\rangle={\rm Tr}A^{\dagger}B$
and related to the H--S distance,
\begin{equation}
D_{\rm HS}(\rho,\sigma)=||\rho-\sigma||_2:=
 \sqrt{{\rm Tr} [(\rho - \sigma)^2]} \ ,
\label{HS1}
\end{equation}
 is Riemannian but not monotone \cite{Oz01}. 
This fact provides an additional 
motivation to investigate other possible metrics and 
distances in the space of mixed quantum states.

Nowadays it is widely accepted that the Bures metric \cite{Bu69,Uh76},
is distinguished by its rather special properties:
it is a Riemannian, monotone metric, which is Fisher--adjusted
\cite{PS96}: in the subspace of diagonal matrices it induces
the statistical distance \cite{BC94}.
Moreover, it is also Fubini--Study--adjusted, since
it agrees with this metric at the space of pure states \cite{Uh95}.
The Bures distance between any two mixed states is a function
of their fidelity \cite{Uh76,Jo94} 
and various properties of this distance are
a subject of considerable interest
(see \cite{Uh92,Hu92,Di93,Hu93,Uh94,Di95,Uh95,Di99}).

A given distance between density matrices
induces a certain measure in this space.
The probability measure in the space of mixed quantum states
induced by the Bures metric was defined by Hall \cite{Ha98}.
He demonstrated that the eigenvectors
of a random state of size $N$
generated according to this measure
are distributed according to the Haar measure on $U(N)$,
derived an explicit expression for the
joint probability density for eigenvalues
of a random state generated according to this measure
and found for $N=2$ the normalization constant $C_N$
for the joint probability density of the eigenvalues.
Further Bures--Hall normalization constants $C_3$
and $C_4$ were calculated by Slater \cite{Sl99b},
but the problem of finding the constants $C_N$ for arbitrary $N$
remained unsolved.

The aim of this work is to compute the volume of the set ${\cal M}_N$ of all
density matrices of size $N$ and the hyperarea of its boundary
according to the Bures measure.
This work is complementary to our parallel paper \cite{ZS03},
in which the volume was computed according to the Hilbert-Schmidt measure,
but the notation and the structure of both papers are not exactly the same.
In this work we treat in parallel two different cases:
of complex density matrices, and its particular subset,
of real density matrices. Both cases are labeled by the
'repulsion exponent' $\beta=2$ for the complex case and
$\beta=1$ for the real case.

The paper is organized as follows. In section II
the Bures distance is defined and its remarkable properties are reviewed.
In section III we derive the Bures measure and express the volume and
the surface of the space of mixed states
as a function of the Hall normalization constants.
Key results of the paper are presented in section IV, in which
we compute the normalization constants $C_N$ for arbitrary $N$, 
see (\ref{Hallconst}),  and calculate explicitly the Bures volume 
of ${\cal M}_N$, see Eq. (\ref{Bvolume}).
In section V we perform a similar analysis of the space of real density matrices.
A concise presentation of the results obtained 
is provided by Eq. (\ref{Sn2explS}),
from which one may read out the
volume of the sets of complex and real density matrices, 
the surface of their boundaries and their 'edges'.

\section{Bures metric}

We shall start the discussion considering the set 
of pure states $|\psi\rangle$ belonging to an $N$--dimensional 
Hilbert space ${\cal H}_N$.
This set has the structure of a complex projective space,
${\mathbb C}P^{N-1}$, and there exists a distinguished,
unitarily invariant measure,  which
corresponds to the geodesic distance on this manifold.
It is called Fubini--Study distance \cite{Fu903,St05}, 
\begin{equation}
D_{\rm FS}(|\varphi_A\rangle,|\varphi_B\rangle) =
 {\rm arccos}(2\kappa-1) = 2{\rm arccos}(\sqrt{\kappa}),
\label{FS}
\end{equation}
where $\kappa=|\langle \varphi_A|\varphi_B\rangle|^2$
denotes the transition probability. This quantity was
generalized for an arbitrary pair of mixed
quantum states by Uhlmann \cite{Uh76}
 \begin{equation}
 F(\rho_A , \rho_B) =
 \Bigl[  \mbox{Tr} \bigl( \sqrt{{\rho}_A}\
{\rho}_B\ \sqrt{{\rho}_A} \bigr) ^{1/2}  \Bigr]^2
= || \sqrt{{\rho}_A} \sqrt{\rho_B} ||_1^2
  \label{fidel2}
\end{equation}
and its remarkable proprieties were studied by Jozsa \cite{Jo94}
who called it {\sl fidelity} \footnote{Although this was the original
definition of Jozsa, some authors use this name for $\sqrt{F}$.}.
It is a symmetric, non-negative, continuous,
concave function of both states and is unitarily invariant.
For any pair of pure states the fidelity reduces to their overlap,
$F={\rm Tr}\rho_A\rho_B=|\langle \psi_A|\psi_B \rangle |^2=\kappa$.
Hence a function of fidelity,
called {\sl Bures length} \cite{Uh95} or {\sl angle} \cite{NC00},
\begin{equation}
D_A(\rho_A,\rho_B) := {\rm arccos}
\sqrt{F (\rho_A,\rho_B)}=\frac{1}{2}{\rm arccos}
\Bigl( 2F(\rho_A,\rho_B)-1\Bigr),
\label{anglefid}
\end{equation}
coincides for any pair of pure states with their
Fubini--Study distance (\ref{FS}).

Fidelity may also be defined \cite{Jo94} as the maximal overlap
between the purifications of two mixed states,
$|\langle \Phi_A|\Phi_B\rangle|^2$, where
$|\Phi_i\rangle \in {\cal H}_{N^2}={\cal H}_N\otimes{\cal H}_{N}$
and both mixed states are obtained by partial tracing
over the auxiliary subsystem $_E$,
$\rho_i={\rm Tr}_E |\Phi_i\rangle \langle \Phi_i|$.
Fidelity allows one to characterize the "closeness" of a pair of
mixed states: it is equal to unity only if both states do coincide.
Another function of fidelity,
\begin{equation}
D_B(\rho_A,\rho_B)=
\bigl(2-2\sqrt{F(\rho_A,\rho_B)}\bigr)^{1/2}=
\sqrt{2-2 {\rm
Tr}\sqrt{\sqrt{\rho_A}\rho_B\sqrt{\rho_A}}}
\label{bures}
\end{equation}
satisfies all axioms of a distance and is called {\sl Bures distance}
\cite{Bu69}. Hence Bures metric is Fubini--Study adjusted,
since for pure states it induces the same geometry.
Several relevant papers on this subject are due to Uhlmann
 \cite{Uh76,Uh92,Uh95}, who showed
that the topological metric of Bures is Riemannian.
 H{\"u}bner provided a detailed analysis of the $N=2$ \cite{Hu92}
and $N=3$ case \cite{Hu93}, while Dittmann
analyzed differential geometric properties of the Bures metric \cite{Di93}.
Relations of the Bures metric to information theory
can be seen in papers by Fuchs and Caves \cite{FC95},
Vedral and Plenio \cite{VP98} while the closely related fidelity
is often used in the comprehensive book by Nielsen and Chang \cite{NC00}. 
It was shown by Braunstein and Caves \cite{BC94}
that for neighbouring density matrices
the Bures distance is proportional to the statistical distance introduced by
Wootters \cite{Wo81} in the
 context of measurements which optimally resolve
neighbouring quantum states.

For diagonal matrices,
$\rho_A={\rm diag}({\vec a})$ and $\rho_B={\rm diag}({\vec b})$,
the expression for the Bures distance simplifies
and becomes a function of the {\sl  Bhattacharyya coefficient}
\cite{Bh43}, $B({\vec a},{\vec b})$,
\begin{equation}
D_B^2(\rho_A,\rho_B)=
2-2  \sum_{i=1}^N \sqrt{ a_i b_i }=
2-2B({\vec a},{\vec b}) \ .
\label{bures2}
\end{equation}
This is the Euclidean distance between two points on a unit circle,
the corresponding vectors form the angle $\theta$ such that
$\cos\theta = {\vec a} \cdot {\vec b}$.
The geodesic distance along the circle
$\vartheta = 2\arccos(1-D_B^2/2)=2D_A$, is called in this case the
{\sl statistical distance},
since it provides a measure of distinguishability
between both probability distributions by statistical sampling.
Another function is called the {\sl Hellinger distance}
 \cite{LV87}
\begin{equation}
D_H({\vec a},{\vec b})=
\sqrt{\sum_{i=1}^N( \sqrt{ a_i}-\sqrt{b_i})^2}=
\sqrt{2-2B({\vec a},{\vec b})} \ .
\label{Heiliger}
\end{equation}

It is clear that for any two diagonal density matrices
their fidelity reduces to the squared Bhattacharyya coefficient,
$F(\rho,\sigma)=B^2({\vec a},{\vec b})$,  while the
Bures distance between both states coincides with
the Hellinger distance between both vectors of eigenvalues,
$D_B(\rho,\sigma)=D_H({\vec a},{\vec b})$.

Let us write the Bures distance for
infinitesimally closed diagonal states
defined by the  vectors $\vec a$ and
 ${\vec a}'={\vec a}+{\rm d}{\vec a}$,
\begin{equation}
{\rm d}s_B^2=\bigl(D_B(\rho_A,\rho_A')\bigl)^2=
2-2 \sum_{i=1}^N a_i \sqrt{1+\frac {{\rm d} a_i}{a_i}} \ .
\label{bures3}
\end{equation}
Since $\sum_{i = 1}^N a_i = 1$ and
      $\sum_{i = 1}^N {\rm d}a_i = 0$, so computing
the Bures line element d$s^2_B$ we need to expand the square
root up to the second order, and obtain
\begin{equation}
{\rm d}s^2_B =  \frac{1}{4} \sum_{i=1}^N
\frac{{\rm d}a_i {\rm d}a_i}{a_i}
= {\rm {Tr}} \bigl( {\rm d} \sqrt{ \rho } \bigr)^2
\label{ds2bur}
\end{equation}
\noindent -- precisely the metric on a sphere, or also,
the statistical metric on the simplex,
also called {\sl Fisher-Rao} metric \cite{Fi25}.
Hence the Bures metric is {\sl Fisher--adjusted} --
restricted to the simplex of diagonal matrices it
agrees with this metric.
For $N=2$ the simplex of diagonal matrices is one-dimensional, and
from the point of view of Bures metric it forms a semicircle.

It is instructive to compare
the geometry induced by the Bures (\ref{bures})
 and the Hilbert-Schmidt metric (\ref{HS1}).
Any $N=2$ density matrix $\rho$ may be expressed
in the Bloch representation
\begin{equation}
\rho=\frac{1}{2} \bigl( {\mathbb I}
  + {\vec \tau} \cdot {\vec \sigma} \bigr)
\label{Bloch}
\end{equation}
where $\vec \sigma$ denotes a vector of three Pauli matrices,
$\{\sigma_x, \sigma_y, \sigma_z\}$, while
the Bloch vector $\vec \tau=\{x,y,z\}$ belongs to
${\mathbb R}^3$.
With respect to the HS metric
the set of mixed states ${\cal M}_2$
forms the Bloch ball $B^3$ with the set of
all pure states at its boundary,
called {\sl Bloch sphere}.
The H--S distance between any mixed states of a qubit is proportional to
the Euclidean distance inside the ball, so the
H--S metric induces the flat geometry in
${\cal M}_2$.
The radius of the Bloch ball depends on the
normalization of the Bloch vector, and
is equal to unity while using the definition (\ref{Bloch}).

\begin{figure} 
   \begin{center}
\includegraphics[width=10.0cm,angle=0]{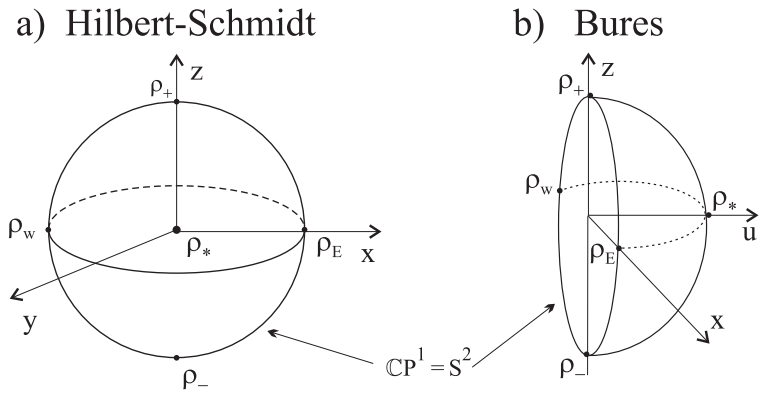}
\caption{Geometry of the set of the mixed states  for $N=2$: 
a) -- the $3$-ball embedded in ${\mathbb R}^3$
induced by the Hilbert-Schmidt distance,
b) -- hemisphere of ${\bf S}^3$ embedded in ${\mathbb R}^4$,
(shown as a cross-section of $R^4$ taken at $y=0$),
 induced by the Bures distance.
The maximally mixed state $\rho_*$
is located at the center of the ball (a),
and at the hyper-pole of ${\bf S}^3$ (b).
The set of pure states is represented by the Bloch 
sphere ${\bf S}^2$ (a)
and by the Uhlmann equator ${\bf S}^2\in {\bf S}^3$ (b).
The cross-section of ${\cal M}_2$ with the plane $z=0$
produces a full circle in the equatorial plane (a),  or a half of a
hyper-meridian joining the states $\rho_W$ and $\rho_E$ (b).}
 \label{fig:puri1}
\end{center}
 \end{figure}

Working in a Cartesian coordinate system we
may rewrite (\ref{Bloch}) as
\begin{equation}
{\rho} = \frac{1}{2}\left( \begin{array}{cc}
1 + z & x - iy \\ x + iy & 1 - z \end{array} \right) \ ,
\hspace{1cm} x^2 + y^2 + z^2 \leq 1 \
\end{equation}
Setting  ${\rho}_A = \rho$, $\rho_B = \rho + {\rm d} \rho$
 in the expression for the Bures distance
(\ref{bures}) and expanding to second order we obtain
one--fourth times the round metric
on the 3-sphere,
\begin{equation}
 {\rm d}s^2_B = \frac{1}{4}\left( dx^2 + dy^2 + dz^2
+ \frac{(xdx + ydy + zdz)^2}{1 - x^2 - y^2 - z^2}\right)
 \ .
\label{burspher}
\end{equation}
Uhlmann noticed this fact in \cite{Uh92} and
suggested embedding of the three dimensional space of the $N=2$
mixed states in ${\mathbb R}^4$. The transformation
\begin{equation}
\rho(x,y,z) \to {\frac{1}{2}} \Bigl( x,y,z,u:=\sqrt{1-x^2-y^2-z^2} \Bigr)
\label{r3r4}
\end{equation}
blows up the $3$-ball into the Uhlmann hyper-hemisphere of 
${\bf S}^3$
of radius $R_2=1/2$.
The maximally mixed state $\rho_*$, mapped into a hyper-pole, is equally
distant from all pure states located at the hyper-equator
- see Fig. \ref{fig:puri1}. The hyper-equator
is equivalent to an ordinary two dimensional sphere and hence isometric
to ${\mathbb C}P^1$.
The comparison between both metrics is very transparent for
real density matrices:
the space of all rebits, which in the HS-metric
has the structure of the $2$-ball, in the Bures metric
has the from of a hemisphere of ${\bf S}^2$.

Since the radius $R_B$ of the Uhlmann hemisphere is equal to $1/2$
we obtain immediately the Bures volume $V$ and the Bures area $S$
for $N=2$
\begin{equation}
V_2^{(2)} =  {\mbox{vol}_B} \bigl( {\cal M}_2  \bigr)  =
\frac{1}{2}S_3 R_B^3=
 \frac {\pi^2}{8}\ ,
\quad \quad
S_2^{(2)} = {\mbox{vol}_B} \bigl( \partial {\cal M}_2 \bigr)  =
  S_2 R_B^2= \pi,
\label{bures22}
\end{equation}
where $S_k$ denotes the volume of the unit sphere ${\bf S}^k$.
This result was obtained by Caves \cite{Ca02},
who also expressed the Bures volume of the set of mixed states 
in form of an integral.
In the further sections of our work we are going
to compute such integrals to derive 
an explicit result for the Bures 
volume for an arbitrary matrix size $N$.

The last property of the Bures metric, we are going to review, concerns
monotonicity. 
 In the classical case of commuting diagonal density
matrices, one  studies
the space of probability vectors ${\vec {a}}=\{a_1,...a_N\}$, where
$a_i\ge 0$ and $\sum_{i=1}^Na_i=1$. This space is isomorphic with an
$N-1$ dimensional simplex $\Delta_{N-1}$.  A discrete dynamics
in this space, ${\vec a}'=T{\vec a}$,
is given by an arbitrary stochastic matrix $T$ of size $N$.
This matrix contains non--negative entries, $T_{ij}\ge 0$,
and the stochasticity condition,
$\sum_{j=1}^N T_{ji}=1$ for $i=1,...,N$ assures that the output vector
${\vec a}'$ is normalized. A distance in the space
$\Delta_{N-1}$ of $N$--dimensional probability vectors is called
{\sl monotone}, if it does not increase under the action of any
stochastic matrix $T$,
\begin{equation}
D_{\rm mon}\bigr({\vec a}, {\vec b} \bigl) \ \ge \
 D_{\rm mon}\bigr(T{\vec a}, T{\vec b} \bigl).
\label{monot1}
\end{equation}
If a geodesic distance in the probability simplex is monotone, 
the corresponding Riemannian  metric is called monotone.
Chentsov showed \cite{Ce82} that every monotone Riemannian
distance in this space is a function of the Fisher statistical distance 
\cite{Fi25}, or the Bhattacharyya coefficient $B$ defined in (\ref{bures2}).

The quantum case is more complicated. Any discrete dynamics
may be described by a quantum operation,
$\Phi:{\cal M}_N \to {\cal M}_N$,
which is trace preserving, Tr$\Phi(\rho)={\rm Tr}\rho=1$,
and completely positive. 
Positivity means that $\Phi(\rho)\ge 0$,
i.e. $\Phi$ sends positive
operators into positive operators.
From a physical point of view it is necessary to assure 
that an operation is well defined on the system coupled to
an ancilla - an auxiliary system which describes the environment.
Hence one requires that $\Phi$ is {\sl completely positive},
which means that any extension of the map 
by an identity operator, $\Phi\otimes {\mathbb I}$,
is positive \cite{OP93}.
Completely positive trace preserving maps play the
role of stochastic matrices $T$ in the quantum case, and are thus called
{\sl stochastic maps}.
In full analogy to (\ref{monot1})
any distance $D$ in the space of mixed quantum states
${\cal M}^{(N)}$ is called {\sl monotone}
if it does not grow under the action of a stochastic map,
\begin{equation}
D_{\rm mon}\bigr(\rho,\sigma \bigl)  \ge
D_{\rm mon}\bigr(\Phi(\rho), \Phi(\sigma) \bigl) \ .
\label{monot2}
\end{equation}
If a monotone distance is geodesic the corresponding metric 
on ${\cal M}_N$ is called monotone.

\begin{figure} 
   \begin{center}
 \vskip -0.2cm
\includegraphics[width=8.0cm,angle=0]{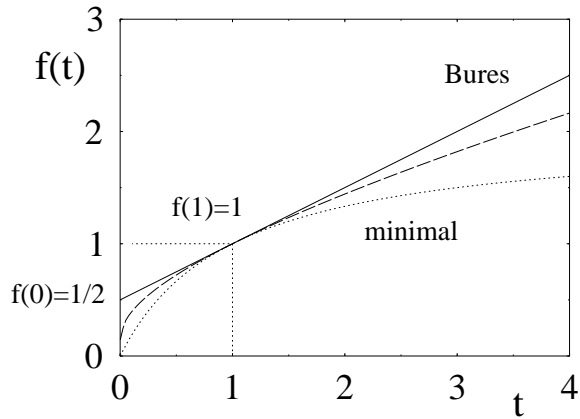}
\vskip -0.2cm
\caption{Morozova--Chentsov functions $f(t)$:
minimal function for maximal metric (dotted line),  
Kubo--Mori metric (dashed line), and maximal function
for the minimal, Bures metric (upper line). These metrics are 
Fisher--adjusted, $f(1)=1$, while Bures metric 
is also Fubini--Study adjusted, $f(0)=1/2$.}
 \label{fig:bur2}
\end{center}
 \end{figure}
In contrast to the classical case, there exist infinitely many
monotone Riemannian metrics on the space of mixed quantum states.
However, as shown by Morozova and Chentsov \cite{MC90}
and Petz and Sud{\'a}r \cite{PS96}, they may be characterized
by a symmetric function $c(x,y)$. In general, the length of a vector
(matrix) $B_{ij}$ may be position dependent;
 at a diagonal point $A={\rm diag }(a_1, a_2, ...,a_N)$
 the squared length of $B$ is given by \cite{PS96}
\begin{equation}
||B||^2_A= \sum_{j=1}^N \frac {B_{jj}}{a_j} +
  2 \sum_{j<k=1}^N c(a_j,a_k) |B_{jk}|^2 \ .
\label{monot3}
\end{equation}
For a diagonal matrix $B$ the second term vanishes and the first
term gives back the unique statistical metric on the simplex.
The off--diagonal elements, which enter the second term in
(\ref{monot3}),  allow for more freedom in the choice of a
monotone Riemannian metric.
The Morozova-Chentsov  function satisfies $c(sx,sy)=s^{-1}c(x,y)$.
The symmetry of this two--argument function $c$ enables to
express it by a function of a single parameter,
$c(x,y)=f(x/y)/y$, such that $f(1/t)=f(t)/t$.

A function $f$ is called {\sl operator monotone}, if 
for any matrices $G$, $H$ the relation 
$0\le G \le H$ implies $0\le f(G) \le f(H)$. 
Here $G\le H$ denotes that their difference is semi positive, 
$H-G \ge 0$. Taking for $G$ and $H$ real numbers we see
that operator monotonicity implies that $f$ maps 
${\mathbb R}_+ \to {\mathbb R}_+$ and is concave. 
The result of Morozova and Chentsov \cite{MC90}, 
completed and extended by Petz and Sud{\'a}r \cite{Pe96,PS96}, 
states that every monotone metric may be written in 
the form (\ref{monot3}) up a proportionality constant. 
Conversely, this equation determines a monotone Riemannian metric, 
if the function $f(t)=1/c(t,1)$ is: a) operator monotone,
and b) self--transposed, $f(1/t)=f(t)/t$. Such functions $f$ are called 
Morozova--Chentsov functions. Comparison of different metrics 
is meaningful under a certain normalization.
Usually one requires: c) $f(1)=1$, 
since for diagonal matrices such a metric 
agrees with the statistical (Fisher) metric, and is called  
{\sl Fisher--adjusted}. 
There exists infinitely many functions $f$
satisfying all three conditions a)--c), 
and leading to Fisher--adjusted metrics \cite{PS96}.
Let us mention three examples here: 
\begin{equation}
f_{\rm min}(t)=\frac{2t}{t+1}, 
{\quad \quad} 
f_{\rm KM}(t)=\frac{t-1}{\ln t}, 
{\quad \quad} 
f_{\rm max}(t)=\frac{1+t}{2}
\label{ft3}
\end{equation}
containing both extremal cases and the 
intermediate Kubo-Mori (Bogolioubov) metric used in quantum statistical 
mechanics \cite{BAR86}
and distinguished by the fact that both affine connections
with respect to this metric are mutually dual \cite{GS01}.
These functions, plotted in Fig. 2,
correspond to 
\begin{equation}
c_{\rm min}(x,y)=\frac{x+y}{2xy}, 
{\quad \quad} 
c_{\rm KM}(x,y)=\frac{\ln x -\ln y}{x-y}, 
{\quad \quad} 
c_{\rm max}(x,y)=\frac{2}{x+y},
\label{ft4}
\end{equation}
so the inverse, $1/c$, is equal to 
the harmonic, logarithmic and arithmetic mean, respectively.
It is known that among all normalised, self-transposed,
operator monotone functions $f$ there is a minimal 
and a maximal one \cite{KA80}. They are
distinguished above by the appropriate labels.
The maximal concave function $f_{\rm max}(t)$, 
the straight line $t/2+1/2$, induces the 
Bures metric and the Bures distance \cite{PS96,LR99}.

If $f(0)=1/2$ the metric induces the natural Riemannian  metric
on the subspace ${\mathbb C}P^{N-1}$ of pure states and is called
{\sl Fubini-Study adjusted} \cite{Uh95,PS96}. The Bures metric is the only
monotone metric which is simultaneously Fisher--adjusted and 
Fubini--Study--adjusted.
This property, valid for any matrix size $N$,
is easiest to discuss for $N=2$.
All metrics defined by (\ref{monot3}) are rotationally
invariant, i.e. the uniform angular distribution 
(of eigenvectors of the density matrix) is independent of the 
radial distribution (of its eigenvalues).
In the one--qubit case the metric depends
only on the radius $r=\sqrt{x^2+y^2+z^2}$ in the unit Bloch ball,
and may be divided into the radial and the normal 
(tangential) components \cite{Pe96},
\begin{equation}
{\rm d} s^2 = \frac{1}{1-r^2} {\rm d} r^2 +
 \frac{1}{1+r}g\Bigl(\frac{1-r}{1+r}\Bigr) {\rm d} n^2
\quad {\rm with \quad} g(t)=\frac{1}{f(t)}
\label{gtft}
\end{equation}
The radial component is independent of the function $f$.
It is easy to see that the value of $f(t)$ at $t=0$
determines the metric at the subspace of pure states,
$r=1$. If $f(0)=0$, then the metric diverges at 
the manifold of pure states, which is the case for 
$f_{\rm KM}$ and $f_{\rm min}$.

The relation between the Morozova-Chentsov function and the metric $g=1/f$
shows that the larger the (normalized) function $f$, the smaller the
metric and the corresponding distance. The Bures metric
given by the {\sl maximal} function $f_{\rm max}(t)$,
is therefore the {\sl minimal} Fisher--adjusted, monotone Riemannian
metric.
These properties single out the Bures metric 
and provide an additional motivation to study the geometry
it induces in the set of mixed quantum states.

\section{Bures measure}

\subsection{Infinitesimal distance}

We are going to compute the Bures distance (\ref{bures})
 between two 
infinitesimally close density matrices of size $N$.
Let us set $\rho_A =\rho ,\ \rho_B = \rho + \delta\rho $:
\begin{equation}
  \sqrt{ \rho_A^{1/2}\rho_B\ \rho_A^{1/2} } = \rho + X + Y  \label{expansion}
\end{equation}
where the matrix $X$ is of order $1$ in $\delta\rho$, 
while $Y$ is of order $2$.
Squaring this equation we obtain to first and second order
\begin{equation}
  \rho^{1/2}\delta\rho\ \rho^{1/2}  = X\rho + \rho X \ ,
 \ \ \ \ \ \ -X^2=Y\rho+\rho Y
\label{1order}
\end{equation}
or in the basis in which $\rho$ is diagonal with eigenvalues $\rho_{\nu}>0$
\begin{equation}
   X_{\nu \mu} =\delta \rho_{\nu \mu}\
\frac{\rho_{\nu}^{1/2} \rho_{\mu}^{1/2}}
{\rho_{\nu} +\rho_{\mu}}\ , \quad \quad
 Y_{\nu\mu}=-(X^2)_{\nu \mu} \frac{1}{\rho_{\nu}+\rho_{\mu}} \ .
\label{Anumu}
\end{equation}
Since ${\rm Tr}\rho=1$, hence ${\rm Tr}\delta\rho=0$
and  ${\rm Tr}X={\rm Tr}\ \delta\rho /2=0$, while
\begin{equation}
 {\rm Tr}Y=-\sum_{\nu\mu}{1\over2\rho_{\nu}}|X_{\nu\mu}|^2=
-\sum_{\nu\mu}{1\over4}{|\delta\rho_{\nu\mu}|^2 \over {\rho_{\nu}+\rho_{\mu}}}  \ .
 \label{TrB}
\end{equation}
Thus we arrive at the result of H{\"u}bner \cite{Hu92} for the Bures metric
\begin{equation}
 ({\rm d}s)_B^2=(D(\rho,\rho+\delta\rho))^2=
{1\over2}\sum_{\nu\mu}{|\delta\rho_{\nu\mu}|^2 \over {\rho_{\nu}+\rho_{\mu}}}    
\label{metricB}\ .
\end{equation}
Note, that if $\rho_{\nu}=0$ and $\rho_{\mu}=0$,
$\delta\rho_{\nu\mu}$ does not appear and therefore
terms where the denominator vanishes have to be excluded.

\subsection{The volume}
Let us rewrite the metric (\ref{metricB}) distinguishing the diagonal elements 
\begin{equation}
 ({\rm d}s)_B^2=\sum_{\nu=1}^N 
 {|\delta\rho_{\nu\nu}|^2 \over {4\rho_{\nu} }}+
 \sum_{\nu<\mu}{|\delta\rho_{\nu\mu}|^2 \over {\rho_{\nu}+\rho_{\mu}}}
    \label{metricB1}\ .
\end{equation}
For diagonal matrices the second term vanishes and the above expression reduces
to (\ref{ds2bur}).
Since ${\rm Tr}\rho=1$
so $\sum_{\nu=1}^N  {\delta\rho_{\nu\nu}}=0$ and not all
$\delta\rho_{\nu\nu}$ are independent. 
Eliminating $\delta\rho_{NN}$ we obtain
\begin{equation}
\sum_{\nu=1}^N  {(\delta\rho_{\nu\nu})^2 \over {\rho_{\nu} }}=\sum_{\nu=1}^{N-1}  
{{(\delta\rho_{\nu\nu})}^2 \over {\rho_{\nu} }} + 
(\sum_{\nu=1}^{N-1}{\delta\rho_{\nu\nu}})^2{1\over\rho_N}
\label{partmetricB1}\ .
\end{equation}
The metric (\ref{partmetricB1}) in the $(N-1)$--dimensional subspace 
is of the form $g_{ik}= (\delta_{ik}/\rho_i +1/\rho_N )$ with 
determinant $ \det g = (\rho_1 +\rho_2 +...+\rho_N )/(\rho_1 \rho_2 ...\rho_N )
=1/ \det \rho$ using  ${\rm Tr}\rho=1$.
 Thus the volume element gains a factor 
$\sqrt{{\rm det} g} =(\det\rho)^{-1/2}$:
\begin{equation}
{\rm d}V_B={1\over 2^{{N^2}-1}}(\det\rho)^{-1/2}\prod_{\nu=1}^{N-1}{\rm d}\rho_{\nu\nu }
\ \prod_{\nu<\mu}{2\over \rho_{\nu}+\rho_{\mu}}\ \prod_{\nu<\mu}
\bigl( \sqrt{2}\ {\rm d}{\rm Re} (\rho_{\nu\mu})\sqrt{2} \ {\rm d}{\rm Im} 
(\rho_{\nu\mu})\bigr)
 \label{dVBures}\ .
\end{equation}
 The total volume of the set ${\cal M}_N$ is given by:
\begin{equation}
V_B={ 2^{1-{N^2}}}\int D\rho\  {\delta( {\rm Tr}\rho-1) 
\over {\det \rho^{1/2}}}\theta(\rho) 
\prod_{\nu<\mu}{2\over  {\rho_{\nu}+\rho_{\mu}}}
 \label{VBures}\
\end{equation}
with $\theta(\rho)=\theta(\rho_1)...\theta(\rho_N)$ a step function that
 restricts integration to nonnegative $\rho$. We introduced the volume element
\begin{equation}
 D\rho = \prod_{\nu=1}^{N}{\rm d}\rho_{\nu\nu }\prod_{\nu<\mu}\bigl( \sqrt{2}\
{\rm dRe} (\rho_{\nu\mu})\sqrt{2} \ {\rm dIm} (\rho_{\nu\mu})\bigr)
  \label{Drho}\ .
\end{equation}
 which differs from the usual definition by a factor $2^{N(N-1)/2}$.
 It follows from the metric for unrestricted Hermitian matrices:
\begin{equation}
 ({\rm d}s)^2=\sum_{\nu=1}^N  { (\delta \rho_{\nu\nu})^2 } 
 + 2\sum_{\nu<\mu}{|\delta \rho_{\nu\mu}|^2}
  \label{metricH}\ .
\end{equation}
Above we considered the case of Hermitian density matrices ($\beta=2$). In
the case of real density matrices ($\beta=1$) there are no imaginary parts
 ${\rm Im} \rho_{\nu\mu}$.

\subsection{Radial coordinates}
Let us diagonalize the matrices $\rho$
\begin{equation}
 \rho = U \Lambda U^{-1}\quad {\rm or}\quad  \rho = O \Lambda O^{-1}\   
     \label{diag}\
\end{equation}
with $U$ unitary for $\beta=2$, $O$ orthogonal for  $\beta=1$, 
$ \Lambda ={\rm diag}(\rho_1, \rho_2,...,\rho_N)$. Then we have
\begin{equation}
 {\rm d}\rho = U [{\rm d}\Lambda +U^{-1}{\rm d}U\Lambda-\Lambda 
 U^{-1}{\rm d}U]U^{-1}\ {\rm or}\ 
 {\rm d}\rho =  O[{\rm d}\Lambda +O^{-1}{\rm d}O\Lambda-\Lambda 
 O^{-1}{\rm d}O]O^{-1} \    
 \label{drho}\ .
\end{equation}
Let us first rewrite the Bures metric in a basis independent way,
\begin{equation}
 ({\rm d}s)_B^2= {1\over 2}\int_0^{\infty}{\rm d}t\ {\rm Tr}
 ({\rm d}\rho {\rm e}^{-\rho t} {\rm d}\rho {\rm e}^{-\rho t})
 \label{basisfree}\ .
\end{equation}
This means it contains left and right multiplications with $\rho$.
Inserting now the expressions for d$\rho$ we see that again we can  
go to the representation where $\rho$ is diagonal and obtain 
metrics in the cases of complex 
density matrices,
\begin{equation}
 ({\rm d}s)^2=\sum_{\nu=1}^N 
\frac{({\rm d}\rho_{\nu})^2} {4\rho_{\nu}} +
 \sum_{\nu<\mu}{(\rho_{\nu}-\rho_{\mu})^2\over
 {\rho_{\nu} + \rho_{\mu}}}|(U^{-1}{\rm d}U)_{\nu\mu}|^2   
   \label{metricBU}\ ,
\end{equation}
and real density matrices  
\begin{equation}
 ({\rm d}s)^2=\sum_{\nu=1}^N  
\frac{ ({\rm d} \rho_{\nu})^2} {4\rho_{\nu}}+ 
\sum_{\nu<\mu}{(\rho_{\nu}-\rho_{\mu})^2\over {\rho_{\nu}+\rho_{\mu}}}
|(O^{-1}{\rm d}O)_{\nu\mu}|^2  
    \label{metricBO}\ .
\end{equation}
In both cases the Bures measure has the following product form
\begin{equation}
{\rm d} V_B^{(\beta)} = {\rm d} \mu^{(\beta)}
 (\rho_1,\rho_2,...,\rho_N) \times {\rm d}
\nu_{\beta}.
  \label{dVdmupr}
\end{equation}
The first factor depends only on the eigenvalues $\rho_i$,  while the
latter, determining the distribution of the eigenvectors of 
the density matrix, is the unique, unitarily invariant, Haar measure on
$U(N)$ for $\beta=2$ or on $O(N)$ for $\beta=1$. 
In the general case of complex density matrices, ($\beta=2)$,
the Bures measure in the simplex of eigenvalues reads
\begin{equation}
P_B(\Lambda)= P_B(\rho_1,\rho_2,...,\rho_N) =
C_N \frac{\delta\bigl(\rho_1+\rho_2+...+\rho_N - 1)}
          {\sqrt{\rho_1\rho_2 \cdots \rho_N}}
   \prod_{\nu<\mu}
\frac{ (\rho_{\nu}-\rho_{\mu})^2 } 
     {\rho_{\nu}+\rho_{\mu}} ,
  \label{mesbur2}
\end{equation}
where the normalization constant $C_N$ assures that 
the integral of $P$ over all eigenvalues equals to unity. 
The measure was derived by Hall \cite{Ha98},
and was later analyzed in \cite{BS01,ZS01}.
To compute the area of the surface of the set of complex density matrices,
and to find the volume of the set of real density matrices we will need 
a family of generalized constants $C_N(\alpha,\beta)$ defined by 
\begin{equation}
 {1\over C_N(\alpha,\beta)}=\int_0^{\infty} \prod_{\nu=1}^{N}
{{\rm d}\rho_{\nu }\over\rho_{\nu}^{1/2}}
\ {\delta(\rho_{1}+...+\rho_N -1) }
\left[\prod_{\nu<\mu}^{1...N}{(\rho_{\nu}-\rho_{\mu})^2\over
\rho_{\nu}+\rho_{\mu}}  \right]^{\beta/2}\prod_{\kappa ={1}}^N
\rho_{\kappa}^{\alpha-1}
 \label{CHall}\ .
\end{equation}
The standard Hall constants entering (\ref{mesbur2})
become a special case of this definition, $C_N=C_N(1,2)$. 
They will be needed to compute the Bures volume of ${\cal M}_N$.

Integrating the Bures measure over the eigenvectors
distributed according to the Haar measure on $U(N)$ or $O(N)$
we obtain
\begin{equation}
 dV_B^{(\beta)}=2^{1-N[1+(N-1)\beta /2]}\det\rho^{-1/2}
\prod_{\nu=1}^{N-1}{\rm d}\rho_{\nu }
\ \prod_{\nu<\mu}\left[{2(\rho_{\nu}-\rho_{\mu})^2\over 
\rho_{\nu}+\rho_{\mu}}\right]^{\beta/2}\ V_{\rm Flag}^{(\beta)}(N)
 \label{dVBuresFlag}\ .
\end{equation}
where
$V_{\rm Flag}^{(\beta)}(N)$ denotes the volume of the flag manifold, 
$Fl_N^{\mathbb C}=U(N)/U(1)^N$ or
$Fl_N^{\mathbb R}=O(N)/O(1)^N$ for $\beta=2,1$.
Although these volumes 
\begin{equation}
V_{\rm Flag}^{(2)}(N)
={\rm Vol}\bigl(Fl_N^{\mathbb C}\bigr)
={(2\pi)^{N(N-1)/2} \over \Gamma(1)\Gamma(2)...\Gamma(N)}
\quad  {\rm and}\quad 
 V_{\rm Flag}^{(1)}(N)=
{\rm Vol} \bigl( Fl_N^{\mathbb R}\bigr)=
{(2\pi)^{N(N-1)/4}\pi ^{N/2} 
\over \Gamma(1/2)\Gamma(2/2)...\Gamma(N/2)}\
\label{VolumeFlag} 
\end{equation}
were computed by Hua several years ago \cite{Hu63},
we provide their short derivation in \cite{ZS03} 
and show that various results for the volumes of the
unitary groups and flag manifolds present in the literature
\cite{Fu01,BST02,TS02,Ca02} are due to different choice of 
normalization.
Formally we may write the volumes in the form:
\begin{equation}
V_{\rm Flag}^{(\beta)}(N)= \prod_{j=1}^N (2\pi)^{(j-1)\beta/2}
 \frac{\Gamma(\beta/2)}{\Gamma(j\beta/2)}
\label{Flagbeta}\ ,
\end{equation}
which shows that it is the product of the volumes
of projective spaces with increasing dimension $j-1$,
real for $\beta=1$, complex for $\beta=2$, and the radius
rescaled by a factor $\sqrt{2}$ coming from the factor $2^{\beta N(N-1)/4}$ --
see appendix of \cite{ZS03}.

Since the unitary or orthogonal rotations produce all permutations,
in (\ref{dVBuresFlag}) we have to restrict to 
a certain order of eigenvalues, $\rho_1< \rho_2<...<\rho_N$.
In order to lift this condition we need to 
divide the result by the number of permutations, equal to $N!$, 
\begin{equation}
V_B^{(\beta)}=2^{1-N[1+(N-1)\beta /2]}\int_0^{\infty}
{{\rm d}\rho_1...{\rm d}\rho_N \over \det\rho^{1/2}} \delta({\rm Tr}\rho-1)
\ \prod_{\nu<\mu}\left[{2(\rho_{\nu}-\rho_{\mu})^2\over 
\rho_{\nu}+\rho_{\mu}}\right]^{\beta/2}\ { V_{\rm Flag}^{(\beta)}(N)\over N!}
 \label{VBuresFlag}\ .
\end{equation}

\subsection{The surface}
The surface consists of the N pieces with one eigenvalue $=0$.
 To calculate the metric on the surface (with $\rho_1=0$) we put 
first ${\rm d}\rho_1=0$ and then $\rho_1=0$ (remember that terms with 
vanishing denominator have to be excluded):
\begin{equation}
 ({\rm d}s)^2=\sum_{\nu=2}^N  {({\rm d}\rho_{\nu})^2 \over {4\rho_{\nu} }}+ 
\sum_{\nu<\mu}^{2...N}{(\rho_{\nu}-\rho_{\mu})^2\over 
{\rho_{\nu}+\rho_{\mu}}}|(U^{-1}{\rm d}U)_{\nu\mu}|^2 +\sum_{\mu=2}^{N} 
\rho_{\mu} |(U^{-1}{\rm d}U)_{1\mu}|^2   
  \label{metricBUS}\ .
\end{equation}
Then with the same reasoning as above the surface element is given by
\begin{equation}
 {\rm d}S=2^{1-(N-1)(1+N\beta/2)} \prod_{\nu=2}^{N-1}
{{\rm d}\rho_{\nu }\over\rho_{\nu}^{1/2}}
\ \left[\prod_{\nu<\mu}^{2...N}{2(\rho_{\nu}-\rho_{\mu})^2\over 
\rho_{\nu}+\rho_{\mu}}
\prod_{\kappa =2}^N2\rho_{\kappa}\right]^{\beta/2}\ V_{\rm Flag}^{(\beta)}(N)
 \label{dSBuresFlag}\
\end{equation}
and the total surface:
\begin{equation}
S_N^{(\beta)}=S_{N,1}^{(\beta)}=
2^{1-(N-1)(1+N\beta/2)} \int \prod_{\nu=2}^{N}{{\rm d}\rho_{\nu }\over\rho_{\nu}^{1/2}}
\ {\delta(\rho_2+...+\rho_N -1)\over(N-1)!} 
\left[\prod_{\nu<\mu}^{2...N}{2(\rho_{\nu}-\rho_{\mu})^2\over \rho_{\nu}+\rho_{\mu}}
\prod_{\kappa =2}^N2\rho_{\kappa}\right]^{\beta/2}\ V_{\rm Flag}^{(\beta)}(N)
 \label{SBuresFlag}\ ,
\end{equation}
where we have taken into account all permutations
of the $N-1$ non--zero eigenvalues.

\subsection{The edges and hyperedges}
There are $N(N-1)/2$ edges which correspond to two vanishing eigenvalues.
Hence there are $(N-2)!$ different permutations of 
positive, generically different eigenvalues which form the edges.
 To calculate the metric on the edges we put first ${\rm d}\rho_1= {\rm d}\rho_2=0$ and then
 $\rho_1= \rho_2=0$:
\begin{equation}
 ({\rm d}s)^2=\sum_{\nu=3}^N  {({\rm d}\rho_{\nu})^2 \over {4\rho_{\nu} }}+ 
\sum_{\nu<\mu}^{3...N}{(\rho_{\nu}-\rho_{\mu})^2\over 
{\rho_{\nu}+\rho_{\mu}}}|(U^{-1}{\rm d}U)_{\nu\mu}|^2 +\sum_{\mu=3}^{N} \rho_{\mu}
(|(U^{-1}{\rm d}U)_{1\mu}|^2 +|(U^{-1}{\rm d}U)_{2\mu}|^2 )   \label{met
ricBUE}\ .
\end{equation}
We see that there do not appear terms $(U^{-1}{\rm d}U)_{\nu\mu}$ 
with $\nu,\mu=1,2$.
Thus the corresponding coset space which has to be
integrated over is the flag manifold $U(N)/[U(2)\times U(1)^{N-2}]$ for $\beta=2$ or
 $O(N)/[O(2)\times O(1)^{N-2}]$ for $\beta=1$.
Since their volume is equal to the ratio
Vol$(Fl_N)/{\rm Vol}(Fl_2)$,
the edge element is then given by
\begin{equation}
 {\rm d}S_{N,2}^{(\beta)}={ 2^{-d_2}} \prod_{\nu=3}^{N-1}
{{\rm d}\rho_{\nu }\over\rho_{\nu}^{1/2}}
\ \left[\prod_{\nu<\mu}^{3...N}{2(\rho_{\nu}-\rho_{\mu})^2\over
\rho_{\nu}+\rho_{\mu}} \prod_{\kappa =3}^N (2\rho_{\kappa})^2
\right]^{\beta/2}\ {V_{\rm Flag}^{(\beta)}(N)\over 
V_{\rm Flag}^{(\beta)}(2)}
 \label{dS2BuresFlag}\
\end{equation}
with the dimension $d_n$
\begin{equation}
d_n= (N-n)(1+(N+n-1)\beta/2)-1
\label{d_n}\ .
\end{equation}
The total area of the edges is
\begin{equation}
 S_{N,2}^{(\beta)}={  2^{-d_2}}\int_0^{\infty} \prod_{\nu=3}^{N}
{{\rm d}\rho_{\nu }\over\rho_{\nu}^{1/2}}
\ {\delta(\rho_3+...+\rho_N-1)\over(N-2)!} \left[\prod_{\nu<\mu}^{3...N}
{2(\rho_{\nu}-\rho_{\mu})^2\over \rho_{\nu}+\rho_{\mu}}
\prod_{\kappa =3}^N(2\rho_{\kappa})^2\right]^{\beta/2}\
{V_{\rm Flag}^{(\beta)}(N)\over V_{\rm Flag}^{(\beta)}(2)}
 \label{S2BuresFlag}\ .
\end{equation}
It is straightforward to generalize this to hyperedges: the 
manifold of states of rank $N-n$ with $n=0,1,...N-1$
has the volume
\begin{equation}
 S^{(\beta)}_{N,n}={ 2^{-d_n}}\int_0^{\infty} \prod_{\nu=n+1}^{N}{{\rm d}\rho_{\nu }\over\rho_{\nu}^{1/2}}
\ {\delta(\rho_{n+1}+...+\rho_N -1)\over(N-n)!} \left[\prod_{\nu<\mu}^{n+1...N}{2(\rho_{\nu}-\rho_{\mu})^2\over \rho_{\nu}+\rho_{\mu}}
 \prod_{\kappa ={n+1}}^N(2\rho_{\kappa})^n\right]^{\beta/2}\
{V_{\rm Flag}^{(\beta)}(N)\over V_{\rm Flag}^{(\beta)}(n)}
 \label{SnBuresFlag}\ .
\end{equation}

\section{Bures volume}

\subsection{Normalization constants}

The above formulae for the volumes of manifolds of states of rank $N-n$
may be expressed in a compact way with use of 
the generalized Hall constants (\ref{CHall}) 
\begin{equation}
S^{(\beta)}_{N,n}={ 2^{-d_n +(N-n)(N+n-1)\beta/4}\over
(N-n)!C_{N-n}(1+n\beta/2,\ \beta)}  \
 {V_{\rm Flag}^{(\beta)}(N)\over V_{\rm Flag}^{(\beta)}(n)}
 \label{SnBuresC}\ .
\end{equation}

We are going to compute the constants $C_N(\alpha,\beta)$ 
for integer $\alpha,\beta$ and 
then continue analytically to general positive $\alpha$ and $\beta$.
Let us go to variables
 $\rho_i=t_i^2$. Then the integral (\ref{CHall})  is
  an integral over the full sphere $R^2=t_1^2+t_2^2+...+t_N^2=1$
 with an homogeneous function in R of order $ \beta N(N-1)/2+2N(\alpha-1)$
as integrand. The radial integral may be replaced by a $\Gamma $ function,
 which amounts to replacing $\delta(R^2-1)$ by ${\rm e}^{-R^2}$.
 The result is the same up to a factor.
Afterwards we may go back to the variables $\rho_i$ and obtain 
\begin{equation}
 {\Gamma[N(2\alpha+(N-1)\beta/2-1)/2]\over C_N(\alpha,\beta)}
=\int_0^{\infty} \prod_{\nu=1}^{N}{{\rm d}\rho_{\nu }\over\rho_{\nu}^{1/2}}
\ {\rm e}^{-(\rho_{1}+...+\rho_N)   } \left[\prod_{\nu<\mu}^{1...N}
{(\rho_{\nu}-\rho_{\mu})^2\over \rho_{\nu}+\rho_{\mu}}
\right]^{\beta/2}\prod_{\kappa ={1}}^N\rho_{\kappa}^{\alpha -1}
 \label{CLaguerre}\ .
\end{equation}

In the next step we want to eliminate the denominators in (\ref{CLaguerre}). 
In this section we discuss the case $\beta=2$ of complex density matrices,
\begin{equation}
{1\over(\rho_1...\rho_N)^{1/2}}\prod_{\nu<\mu}{1\over\rho_{\nu}+\rho_{\mu}}=
\int{\rm e}^{-\sum\rho_i X_{ii}^2-\sum_{i<j}|X_{ij}|^2(\rho_i+\rho_j)}
 \prod_{i=1}^{N}{{\rm d}X_{ii}\over \sqrt{\pi}}\prod_{i<j}
{{\rm d \ Re}X_{ij}~ {\rm d\ Im} X_{ij}\over \pi}
\label{Denominator}
\end{equation}
Using the volume element $DX$ of a Hermitian Matrix X
 (\ref{Drho}) we may rewrite this as
\begin{equation}
{1\over(\rho_1...\rho_N)^{1/2}}\prod_{\nu<\mu}{1\over\rho_{\nu}+\rho_{\mu}}=
\pi^{-N^2/2}\ 2^{-N(N-1)/2}\ \int DX\ {\rm e}^{-{\rm Tr}\ \rho X^2}\ .
\label{randomM}
\end{equation}
This expression is basis independent and provides a way 
to describe the Bures probability distribution in a random matrix description. 
With this we may write for $\beta=2$
\begin{equation}
 {\Gamma(N^2/2+N(\alpha-1))\over C_N(\alpha,2)}= \pi^{-N^2/2}\ 2^{-N(N-1)/2}
 {N!\over V_{\rm Flag}^{(2)}(N)}\ \int D\rho\ \theta(\rho)\det\rho^{\alpha-1}\
\int DX\ {\rm e}^{-{\rm Tr}\ \rho(1+X^2)}
 \label{CLaguerre2}\ .
\end{equation}
Now we use a formula which is an extension of the Ingham-Siegel integral
 \cite{Fy02}
\begin{equation}
\left[\det\left({\delta\over \delta\rho}\ +\epsilon\right)\right]^{-N}\ 
\delta(\rho)= A_N\theta(\rho)\ {\rm e}^{-{\rm Tr}\ \epsilon\rho}
\label{HSformula}\
\end{equation}
valid for a positive Hermitian matrix $\epsilon$ and with a constant $A_N$.
 Let us derive this formula:
 \begin{equation}
\left[\det\left({\delta\over \delta\rho}\ +\epsilon\right)\right]^{-N}\ 
\delta(\rho)=  \pi^{-N^2}\int {\rm d}\psi_1...{\rm d}\psi_N\ {\rm e}^{-[\psi_1^{\dagger} 
(\delta/\delta\rho +\epsilon)\psi_1+...+\psi_N^{\dagger}
(\delta/\delta\rho +\epsilon)\psi_N]}\ \delta(\rho)
\label{HSformula1}\
\end{equation}
with $N$-dimensional complex vectors $\psi_1,...,\psi_N$.
The translation operator shifts the argument $\rho$ of the $\delta$--function
by the matrix $\psi_1\otimes\psi_1^{\dagger}+...+\psi_N\otimes\psi_N^{\dagger}$.
 Then the integration can be done and yields
$$
\left[\det\left({\delta\over \delta\rho}\ +\epsilon\right)\right]^{-N}\ \delta(\rho)=
{\rm e}^{-{\rm Tr}\ \epsilon\rho}\ \pi^{-N^2}\ \int {\rm d}\psi_1...{\rm d}\psi_N\
\delta(\rho-(\psi_1\otimes\psi_1^{\dagger}+...+\psi_N\otimes\psi_N^{\dagger}))\ .$$
For $\rho>0$ we may rescale $\psi \to \sqrt{\rho}\psi$. 
The Jacobian cancels out and the result is equation (\ref{HSformula}).
 The constant $A_N$ can be calculated from integrating this equation
 (\ref{HSformula}) over $\rho$. With partial integration we find

$${2^{N(N-1)/2}\over \det\epsilon^N}= 
A_N\ \int D\rho\ \theta(\rho)\ {\rm e}^{-{\rm Tr}\ \epsilon\rho}\ .$$
This is valid for any Hermitian matrix $\epsilon>0$.
The coefficient $A_N$ is independent of the matrix $\epsilon$.
To calculate $A_N$ we put $\epsilon= {\mathbb I}$.
The integral can again be done using the Laguerre ensemble:
\begin{equation}
{1\over A_N}={V_{\rm Flag}^{(2)}(N)\over 2^{N(N-1)/2}N!}\
\prod_{j=1}^N\ \Gamma(1+j)\Gamma(j)=\pi^{N(N-1)/2}\ \Gamma(1)...\Gamma(N)
\label{1/AN}\ .
\end{equation}

\subsection{Main results}
Using equations (\ref{CLaguerre2}, \ref{HSformula}, \ref{1/AN}) we may write
\begin{equation}
 {\Gamma(N^2/2+N(\alpha-1))\over C_N(\alpha,2)} =
 {\pi^{-N^2/2}\ 2^{-N(N-1)/2}\over A_N} 
{N!\over V_{\rm Flag}^{(2)}(N)}\ \int D\rho\int DA\  
 \det\rho^{\alpha-1}\ \left[\det\left({\delta\over \delta\rho} 
+ 1 + A^2\right)\right]^{-N}\ \delta(\rho)
 \label{CN(a,2)}\ .
\end{equation}
For $\alpha=1$ we obtain immediately by partial integration:
\begin{equation}
 {\Gamma(N^2/2)\over C_N(1,2)}= {\pi^{-N^2/2}N!\over A_N 
V_{\rm Flag}^{(2)}(N)}\ \int {DA\over   \det(1+A^2)^{N}}\ 
 = {\pi^{-N^2/2}\over A_N} \int\prod_{i=1}^N {{\rm d}a_i\over (1+a_i^2)^N}\ 
\prod_{i<j}(a_i-a_j)^2
 \label{CN(1,2)}\ .
\end{equation}
The last integral may be calculated in the complex plane and yields:
 $(2\pi)^NN!2^{-N^2}$, 
such that the Bures--Hall constants are given by
\begin{equation}
  C_N=C_N(1,2)=  {2^{N^2-N}\ \Gamma(N^2/2)\over \pi^{N/2}\ \Gamma(1)...\Gamma(N+1)}
 \label{Hallconst}\ .
\end{equation}
This coincides with the known values for the Bures--Hall constants,
$C_2=2/\pi$, $C_3=35/\pi$ and $C_4=2^{11} 35/\pi^2$ 
and confirms some conjectures of Slater concerning $C_5$ and $C_6$ \cite{Sl99b}.  
The Bures volume of the set of complex density matrices of size $N$
is now given by
\begin{equation}
  V_N^{(2)}=  {2^{1-N(N+1)/2 }\over C_N(1,2)}\ {V_{\rm Flag}^{(2)}(N)\over N!}=
2^{1-N^2} {\pi^{N^2/2}\over \Gamma(N^2/2)}
 \label{Bvolume}
\end{equation}
and for $N=2$ reduces to (\ref{bures22}).
Note that the factor $2^{1-N^2}$ enters from the beginning in the definition.
 It disappears, if we increase d$s$ (\ref{metricB1}) by a factor of $2$.
 The Bures volume is exactly equal to the $(N^2-1)$-dimensional 
volume of an $(N^2-1)$ dimensional hemisphere with radius $1/2$.
Equations (\ref{Hallconst}) and (\ref{Bvolume})
constitute the central results of this work.

\subsection{Area of surface and edges}
For the surface and the edges of the Bures manifold 
we need $\alpha \ne 1$. Scaling $ \rho \to \epsilon^{-1/2}\ 
\rho\ \epsilon^{-1/2} $  for $ \epsilon $ any Hermitian matrix
 $>0$ shows that the integral (\ref{CN(a,2)}) can be obtained from
\begin{equation}
\int D\rho \det\rho^{\alpha-1}\ \left[\det\left({\delta\over \delta\rho} 
+\epsilon\right)\right]^{-N}\ \delta(\rho)=B_N\ \det\epsilon^{1-N-\alpha}
\label{BNalpha}\
\end{equation}
with $B_N$ independent of $\epsilon$. To compute $B_N$ we put $\epsilon= {\mathbb I}$ 
and go back using equation (\ref{HSformula}). This again leads to an 
integral over the Laguerre ensemble and we obtain:
\begin{equation}
 B_N=A_N\ {V_{\rm Flag}^{(2)}(N)\over N!}\ 
\prod_{j=1}^N\Gamma(1+j)\Gamma(\alpha+j-1)=2^{N(N-1)/2} 
\prod_{j=1}^N{\Gamma(\alpha+j-1)\over \Gamma(j)}
\label{B_N}\ .
\end{equation}
Such an integral appeared already for the H--S metric \cite{ZS03}.
 Thus we have
\begin{eqnarray}
 {\Gamma[N^2/2+N(\alpha-1)]\over C_N(\alpha,2)}= {\pi^{-N^2/2} N!
  \over A_N \    V_{\rm Flag}^{(2)}(N)}\
\int{ DX  \over \det(1+X^2)^{N+\alpha-1}} \nonumber\\
=\prod_{j=1}^N{\Gamma(\alpha+j-1)\over\Gamma(1/2)}\
\int_{-\infty}^{+\infty}\prod_{i=1}^{N}{{\rm d}x_i\over
(1+x_i^2)^{N-1 +\alpha}}\ \prod_{j<k}(x_j-x_k)^{2}
 \label{CN(a,2)1}\ .
\end{eqnarray}
Thus the ensemble is reduced to a kind of the Lorentz ensemble.
The last integral can be found in the book by Mehta \cite{Me91}, (page 350)
 following from the Selberg's integral:
\begin{equation}
 J(1,1,N+\alpha-1,N+\alpha-1,1,N)=\pi^N 2^{-N^2+N-2N(\alpha-1)}
\prod_{j=1}^N{\Gamma(1+j)\Gamma(N+2\alpha-1-j)\over \Gamma(N+\alpha-j)^2}
 \label{JNalpha}\ .
\end{equation}
As a result we obtain
\begin{eqnarray}
{\Gamma(N^2/2+N(\alpha-1))\over C_N(\alpha,2)}=
\pi^{N/2} 2^{-N^2+N-2N(\alpha-1)} \prod_{j=1}^N
{\Gamma(1+j)\Gamma(2\alpha-2+j)\over \Gamma(\alpha+j-1)}
   \label{CN(a,2)2}\ .
\end{eqnarray}
From equation (\ref{SnBuresC}) we obtain for the volumes
of submanifolds of rank $N-n$ of the space of complex density matrices
\begin{equation}
 S^{(2)}_{N,n}={2^{-d_n+(N-n)(N+n-1)/2}\over 
C_{N-n}(1+n,2)(N-n)!}\ {V_{\rm Flag}^{(2)}(N)\over V_{\rm Flag}^{(2)}(n)}
\label{Sn2}\ .
\end{equation}
The result is rather simple \footnote{An error in this formula
which appeared in print, 
{\sl J. Phys.} {\bf A 36}, 10083-10100 (2003), 
is corrected here. We are thankful to P. Slater
for drawing our attention to this issue.}
\begin{equation}
%
S^{(2)}_{N,n}=2^{-d_n}{\pi^{(d_n+1)/2}\over 
\Gamma[(d_n+1)/2]} 
\prod_{j=0}^{N-n-1} \frac{j! (j+2n)!}{[(j+n)!]^2}
 \label{Sn2expl} 
\end{equation}
and reduces to (\ref{Bvolume}) for $n=0$.
Here $d_n= N^2-n^2 -1$ denotes the dimensionality of each submanifold. 
%
Observe that the above prefactor equals to 
the volume of a $d_n$--dimensional hyper-hemisphere with radius $1/2$. 
The above results allows us to find the 
ratio between the surface and the volume, 
\begin{equation}
\gamma_B= {S^{(2)}_{N,1}\over S^{(2)}_{N,0}}={ 2\over \sqrt{\pi} }
{\Gamma(N^2/2)\over \Gamma(N^2/2-1/2)}\ N \ .
\label{S/V}\ 
\end{equation}
It behaves as $N^2$
and in the limit of large matrix size it
grows proportionally to the dimensionality $N^2-1$ of the space
of mixed states. Note that such a ratio for the
hemispheres of dimensionality $N^2$ is proportional to $N$.
For pure states $n=N-1$ we obtain the volume
\begin{equation}
   S^{(2)}_{N,N-1}= {\rm Vol}({\mathbb C}P^{N-1})= {\pi^{N-1} \over (N-1)!}
\label{S^N-1}\ ,
\end{equation}
which is equal to the volume of a $2(N-1)$ dimensional real ball of 
radius $1$ and this equals the volume of a $(N-1)$--dimensional
 complex projective space, which is a 
correct consequence of the Bures metric for pure states. 
By the way, up to a scale factor this is the same for the
 Hilbert-Schmidt metric \cite{ZS03}. 
Differences come only for mixed states.

\section{Real density matrices \& Lorentz type ensemble}

To compute the volume of the set of real density matrices we need
to find the generalized Hall constants (\ref{CHall}) for $\beta=1$.
To do so we write
\begin{equation}
{1\over(\rho_1...\rho_N)^{1/2}}\prod_{\nu<\mu}
{1\over\ (\rho_{\nu}+\rho_{\mu})^{1/2}}=
\int{\rm e}^{-\sum\rho_i X_{ii}^2-\sum_{i<j}X_{ij}^2(\rho_i+\rho_j)}
 \prod_{i=1}^{N}{{\rm d}X_{ii}\over \sqrt{\pi}}\prod_{i<j}
{{\rm d} X_{ij}  \over \sqrt{\pi}}
\label{Denominator1}
\end{equation}
Let us introduce the volume element of a real symmetric matrix $X$ with the metric
\begin{equation}
 ({\rm d}s)^2= {\rm Tr}({\rm d}X)^2=\sum_{i=1}^N 
 {({\rm d}X_{ii})^2 }  + 2\sum_{i<j}{ {\rm d}X_{ij}^2}
    \label{metricS}\ ,
\end{equation}
\begin{equation}
  DX= \prod_i {\rm d}X_{ii}\ \prod_{j<k}({\rm d}X_{jk}\sqrt{2}).
\end{equation}
Then we have
\begin{equation}
{1\over(\rho_1...\rho_N)^{1/2}}\prod_{\nu<\mu}
{1\over (\rho_{\nu}+\rho_{\mu})^{1/2}}=
\pi^{-N(N+1)/4}\ 2^{-N(N-1)/4}\ \int DX\ {\rm e}^{-{\rm Tr}\ \rho X^2}\ .
\label{randomMS}
\end{equation}
With this we may write the formula for $\beta=1$
\begin{equation}
 {\Gamma(N(\alpha-1/2+(N-1)/2))\over C_N(\alpha,1)}=  (2\pi)^{-N(N+1)/4}\ 
2^{N/2} \int_0^{\infty} {\rm d}\rho_1...{\rm d}\rho_N \prod_{\nu<\mu}
|\rho_{\nu}-\rho_{\mu}|^{1}\det\rho^{\alpha-1}\int DA\ 
{\rm e}^{-{\rm Tr}(\rho(1+A^2)}
 \label{CLaguerre2S}\ .
\end{equation}
Now with $\epsilon$ a real symmetric positive matrix we want to calculate 
for real symmetric $\rho$:
\begin{eqnarray}
\left[\det\left({\delta\over \delta\rho}\ +\epsilon\right)\right]^{-N/2}\
 \delta(\rho)=  \pi^{-N^2/2}\int {\rm d}\psi_1...{\rm d}\psi_N\ 
{\rm e}^{-[\psi_1^{\top}
 (\delta/\delta\rho +\epsilon)\psi_1+...+\psi_N^{\top}(\delta/\delta\rho +\epsilon)\psi_N]}\ \delta(\rho)
\nonumber\\
=  {\rm e}^{-{\rm Tr}\ \epsilon\rho}\ \pi^{-N^2}\
 \int {\rm d}\psi_1...{\rm d}\psi_N\ \delta(\rho-(\psi_1\otimes
\psi_1^{\top}+...+\psi_N\otimes\psi_N^{\top}))
\label{HSformula2}\ ,
\end{eqnarray}
with real vectors $\psi_1,...,\psi_N$. For $\rho>0$ we may rescale $\psi \to \sqrt{\rho}\psi$,
 then the volume element is $\propto \det\rho^{N/2}$ and the 
$\delta$-function $\propto \det\rho^{-(N+1)/2}$. This implies
\begin{equation}
\left[\det\left({\delta\over \delta\rho}\ 
+\epsilon\right)\right]^{-N/2}\ \delta(\rho)= 
K_N\ 
\theta(\rho)\ {\rm e}^{-{\rm Tr}\ \epsilon\rho}\det\rho^{-1/2}
\label{HSformulaS}\
\end{equation}
with $K_N$ a constant independent of $\epsilon$. With partial integration we find
$${2^{N(N-1)/4}\over \det\epsilon^{N/2}}=  K_N
\ \int D\rho\ \theta(\rho)\ {\rm e}^{-{\rm Tr}\ 
\epsilon\rho}\ \det\rho^{-1/2}\ .$$
To calculate $K_N$ we put $\epsilon={\mathbb I}$ and
 find again with the help of 
the Laguerre ensemble \cite{Me91} (page 354)
\begin{equation}
{1\over K_N}=
{V_{\rm Flag}^{(1)}(N)\over 2^{N(N-1)/4}N!}\ \prod_{j=1}^N\ 
{\Gamma(1+j\beta/2)\Gamma[1/2+(j-1)\beta/2]\over \Gamma(1+\beta/2)}
\label{1/CN}\ .
\end{equation}

To obtain the generalized normalization constants for $\beta=1$
and arbitrary $\alpha$  we may write
\begin{equation}
\frac {\Gamma\bigl[N\bigl(2\alpha-1+(N-1)/2\bigr) /2\bigr]} 
{C_N(\alpha,1)} =
{ (2\pi)^{-N(N-1)/4}2^{N/2}N!\over K_N
  V_{\rm Flag}^{(1)}(N)}\ \int D\rho\int DX\
   \det\rho^{\alpha-1/2}\ \left[\det\left({\delta\over \delta\rho} +1+X^2\right)
\right]^{-N/2}\ \delta(\rho)
 \label{CN(a,beta)S}\ .
\end{equation}
Scaling $ \rho \to \epsilon^{-1/2}\ \rho\ \epsilon^{-1/2} $  for 
$\epsilon$ any real symmetric matrix $>0$ shows the identity
\begin{equation}
\int D\rho \det\rho^{\alpha-1}\
\left[\det\left({\delta\over \delta\rho} +\epsilon\right)\right]^{-N/2}\
\delta (\rho) =L_N\ \det\epsilon^{(1-N-2\alpha)/2}
\label{DNalpha}\
\end{equation}
with $L_N$ a constant independent of the matrix $\epsilon$.
To compute $L_N$ we put $\epsilon={\mathbb I}$ and go back.
 Again we obtain an integral over the Laguerre ensemble with the result:
\begin{equation}
L_N= 2^{N(N-1)/4} \prod_{j=1}^N{\Gamma[\alpha+(j-1)/2]\over
\Gamma[1/2+(j-1)/2]}
\label{D_NS}\ .
\end{equation}
This formula is very similar to (\ref{B_N})
for the constant $B_N$ in the complex case.
In fact it is not difficult to rewrite both results
in one expression as a function of $\beta$.
This allows us to write down an equation,
which is valid as well in the complex or in the real case,
for $\beta=2$ and $\beta=1$, respectively, 
\begin{eqnarray}
 \frac{\Gamma\bigl[N\bigl(2\alpha-1+(N-1)\beta/2\bigl)/2\bigr]}
 {C_N(\alpha,\beta)}
 =\prod_{j=1}^N{\Gamma[\alpha+(j-1)\beta/2]\over\Gamma(1/2)}\
\int_{-\infty}^{+\infty}\prod_{i=1}^{N}
{{\rm d}x_i\over (1+x_i^2)^{(N-1)\beta/2+\alpha}}\
 \prod_{j<k}|x_j-x_k|^{\beta}
 \label{CN(a,beta)1S}\ .
\end{eqnarray}
As result we have a Lorentz type ensemble.
   The last integral can be found in the book by Mehta (page 350) with
 $\gamma=(N-1)\beta/2+\alpha$:
\begin{equation}
 J(1,1,\gamma,\gamma,\beta/2,N)=
\pi^N 2^{N((N-1)\beta/2+2-2\gamma)}
 \prod_{j=0}^{N-1}{\Gamma[1+(j+1)\beta/2]\
 \Gamma[2\gamma-(N+j-1)\beta/2-1]\over \Gamma(1+\beta/2)\
 [\Gamma(\gamma-j\beta/2)]^2}
 \label{JNalphaS}\ .
\end{equation}
Inserting this into equ.(\ref{CN(a,beta)1S}) we obtain
\begin{eqnarray}
\frac {\Gamma\bigl[N\bigl(2\alpha-1+(N-1)\beta/2\bigr)/2\bigr]}
 {C_N(\alpha,\beta)} =
{\pi^{N/2}\over 2^{N((N-1)\beta/2+2(\alpha-1))}} \prod_{j=1}^N{\Gamma(1+j\beta/2)\
 \Gamma[(N-j)\beta/2+2\alpha-1]\over \Gamma(1+\beta/2)\  
\Gamma[(N-j)\beta/2 +\alpha]}
   \label{CN(a,2)2S} \ .
\end{eqnarray}
From the above equality expression we may read out the explicit form of 
the generalized Hall constants $C_N(\alpha,\beta)$. 
They allow us to obtain a fairly general expression
for the Bures volume of the submanifold of the states of rank 
$N-n$ of the set of complex ($\beta=2$)
or real ($\beta=1$) density matrices    
\begin{equation}
 S^{(\beta)}_{N,n}=2^{-d_n}{\pi^{(d_n+1)/2}\over 
\Gamma((d_n+1)/2)}
\prod_{j=1}^{N-n}{\Gamma(j\beta/2)\
\Gamma[1+(2n+j-1)\beta/2]\over\Gamma[(n+j)\beta/2]\ 
\Gamma[1+(n+j-1)\beta/2]}
\label{Sn2explS}\
\end{equation}
where $d_n=(N-n)[1+(N+n-1)\beta/2]-1$ represents 
the dimensionality of the manifold.
In a sense, this formula may be considered as a brief
summary of the paper:
for $n=0$ the last factor simply equals unity
and (\ref{Sn2explS}) gives the Bures volume of the entire 
space of density matrices, equal to that of a $d_0$-dimensional 
hyper--hemisphere
with radius $1/2$. In the case $n=1$ 
we obtain the volume of the surface of this set,
while for $n=N-1$ we get the volume of the set 
of pure states
\begin{equation}
   S^{(\beta)}_{N,N-1}= {\pi^{(N-1)\beta/2}
\Gamma(\beta/2) \over \Gamma(N\beta/2)}
\label{S^N-1S}\ ,
\end{equation}
which for $\beta=1(2)$ gives correctly the volume of the real
(complex) projective space of dimension $N-1$. This
result can be seen directly from equations
(\ref{SnBuresFlag},\ref{Flagbeta}). In the case of a real projective
space the volume is equal to that of a hemisphere of
dimension $N-1$ and radius $1$: $\pi^{N/2}/\Gamma(N/2)$. The final
 formulas are probably correctly continued to general positive $\beta$,
though we have strictly proven them only for $\beta=1,2$.


\section{Concluding remarks}

In this work we have computed the volume of the set of 
complex (real) density matrices with respect to the Bures measure.
Interestingly, this volume is equal to that of
a hyper-hemisphere of radius $1/2$ and the same dimensionality.
However, this fact does not mean that the geometry
of the set of complex density matrices induced by the Bures metric 
is identical with the geometry of a hemisphere of dimensionality $N^2-1$
embedded in ${\mathbb R}^{N^2}$.  This is the case for $N=2$,
for which the set ${\cal M}_2$ is isomorphic with the Uhlmann
hemisphere  $\frac{1}{2} {\bf S}^3$, 
of a constant constant positive curvature,  embedded in ${\mathbb R}^4$.

On the other hand, in the general case,  $N \ge 3$,
the curvature of the set ${\cal M}_2$ 
induced by the Bures metric is not constant. 
To explain this fact, rigorously analyzed by Dittmann \cite{Di99b}, 
consider the set of pure states for $N=3$. 
This manifold forms a complex projective space,
${\mathbb C}P^2$,  and contains, on one hand,
the sphere ${\bf S}^2={\mathbb C}P^1$ and on the other, the real projective space
${\mathbb R}P^2$. The latter is equivalent to a hemisphere,
(with all antipodal points identified),
so it has a different geodesic length and
curvature different than ${\mathbb C}P^1$. Hence the set of mixed states
for $N>2$ analyzed with respect to the Bures metric,
which preserves the geometry of the manifold of pure states,
cannot have a constant curvature.
Note that the Hilbert--Schmidt distance induces a flat
geometry of the space of mixed states,
(i.e. with the curvature equal to zero), 
for an arbitrary dimensionality $N$.

The Bures measure, related to statistical distance,
Fisher information and Jeffreys' prior, 
is often used to define and compute the probability
that a random density matrix
satisfies a certain condition. For instance
such an approach was applied to find the probability 
that a random state acting on a given
composed Hilbert space is separable \cite{Sl99,Sl02,Sl03}. 
In this work we computed the Bures-Hall normalization constants,
which may be useful to extend such calculations
for higher--dimensional problems.

Results presented in this work may also be considered 
as a contribution to the theory of random matrices. 
A closest work in this field is due to 
Eynard and Kristjansen \cite{EK95}, who 
analyzed the O(n) model on a random lattice.
We introduced a different class of
ensembles of random matrices
 defined by the joint probability density of eigenvalues
\begin{equation}
 P^{(\beta)}(\rho_1,\rho_2,...,\rho_N) = C_N(1,\beta) 
\frac{\delta\bigl( \rho_1+\rho_2+...+\rho_N -1)}
          {\sqrt{\rho_1\rho_2 \cdots \rho_N}}
   \prod_{\nu<\mu}
\frac{ | \rho_{\nu}-\rho_{\mu}|^{\beta}  } 
     {( \rho_{\nu}+\rho_{\mu})^{\beta/2}  } ,
  \label{genbus}
\end{equation}
with normalization constants computed in (\ref{CN(a,2)2S}),
which for $\beta=2$ and $\beta=1$ 
gives the Bures measure on the set of complex and real 
density matrices, respectively. The density of eigenvalues
for these ensembles will be studied in a forthcoming publication
\cite{SZ04}.
It might be also instructive to analyze a family of such
ensembles by allowing the parameter $\beta$ to admit
arbitrary real values. In the limit $\beta \to 0$ 
this  distribution tends to the Dirichlet distribution 
of order $1/2$. It describes the 
distribution of squared components of a real random vector,
i.e. a column (or a row) of a random ortogonal matrix
generated according to the Haar measure on $O(N)$ \cite{Zy99,ZS01}.


It is a pleasure to thank P. Slater for inspiring correspondence
and many helpful remarks. We are also grateful to  
U. Ambresch, I. Bengtsson and W. S{\l}omczy{\'n}ski 
for fruitful discussions.
Financial support by Komitet Bada{\'n} Naukowych in Warsaw under
the grant 2P03B-072~19
and the Sonderforschungsbereich /Transregio 12
der Deutschen Forschungsgemeinschaft 
is gratefully acknowledged.


\begin{thebibliography}{99}

\bibitem{PS96} Petz D and Sud\'{a}r C 1996
{\sl J. Math. Phys.} {\bf 37} 2662

\bibitem{ZSlo01} {\.Z}yczkowski K and S{\l}omczy{\'n}ski W  2001
 {\sl J. Phys.} {\bf A 34} 6689

\bibitem{Ru94} Ruskai M B 1994
 {\sl Rev. Math. Phys.} {\bf 5} 1147 


\bibitem{Oz01} Ozawa M 2001
    {\sl Phys. Lett.} {\bf A 268} 158

\bibitem{Bu69}  Bures  D J C 1969 {\sl Trans. Am. Math. Soc.}
  {\bf 135}, 199

\bibitem{Uh76}   Uhlmann A  1976 {\sl Rep. Math. Phys.} {\bf 9} 273

\bibitem{BC94} Braunstein S L and Caves C M 1994
{\sl Phys. Rev. Lett.} {\bf 72} 3439

\bibitem{Uh95} Uhlmann A 1995
{\sl Rep. Math. Phys.} {\bf 36} 461

\bibitem{Jo94} Jozsa R 1994
 {\sl J. Mod. Opt.} {\bf 41} 2315 

\bibitem{Uh92}  Uhlmann A  1992
in {\sl Groups and related Topics}  Gierelak R et. al.
(eds.), (Dodrecht:  Kluver)

\bibitem{Hu92} H{\"u}bner M 1992
{\sl Phys. Lett.} {\bf A 163} 239

\bibitem{Di93} Dittmann J 1993
 {\sl Sem. S. Lie} {\bf 3} 73

\bibitem{Hu93} H{\"u}bner M 1993
{\sl Phys. Lett.} {\bf A 179} 226

\bibitem{Uh94} Uhlmann A 1996
{\sl J. Geom. Phys.} {\bf 18} 76

\bibitem{Di95} Dittmann J 1995
{\sl Rep. Math. Phys.} {\bf 36} 309

\bibitem{Di99} Dittmann J 1999
{\sl J. Phys.} {\bf A 32} 2663

\bibitem{Ha98}  Hall M J W 1998 {\sl Phys. Lett.}
{\bf A 242} 123

\bibitem{Sl99b}  Slater P B 1999
{\sl J. Phys.} {\bf A 32} 8231

\bibitem{ZS03}  {\.Z}yczkowski K  and  Sommers H.-J. 2003 
{\sl J. Phys.} {\bf A 36} 10115 

\bibitem{Fu903}  Fubini G 1903 
{\sl Atti Instituto Veneto} {\bf 6} 501

\bibitem{St05} Study E 1905 
  {\sl Math. Annalen} {\bf 60} 321

\bibitem{NC00} Nielsen  M A and Chuang I L 2000,
{\sl Quantum Computation and Quantum Information}
 (Cambridge: Cambridge University Press)

\bibitem{FC95} Fuchs C A and Caves C M 1995
{\sl Open Sys. Inf. Dyn} {\bf 3} 1 

\bibitem{VP98} Vedral V and Plenio M B 1998
{\sl  Phys. Rev.} {\bf A 57} 1619

\bibitem{Wo81} Wootters W K 1981
{\sl Phys. Rev.} {\bf D 23} 357

\bibitem{Bh43}  Bhattacharyya A 1943
 On a measure of divergence between two statistical 
populations defined by their probability distributions
  {\sl Bull. Calcutta Math. Soc.} {\bf 35} 99

\bibitem{LV87} F. Liese and Vajda I 1987
  {\sl Convex Statistical Distances}
  (Leipzig: Teubner)

\bibitem{Fi25} Fisher R A 1925
{\sl Proc. Cambridge Philos. Soc.} {\bf 22} 700

\bibitem{Ca02} Caves C 2002 unpublished notes
"Measures and volumes for spheres, the probability simplex,
projective Hilbert space and density operators";
see http://info.phys.unm.edu/~caves/reports/reports.html

\bibitem{Ce82} Cencov N N 1982 {\sl Statistical decision rules
 and optimal inference} Translation of Math. Monogr. 53 
(Providence: American Mathematical Society) 

\bibitem{OP93} Ohya M and Petz D 1993
  {\sl Quantum Entropy and Its Use}
 (Berlin: Springer)

\bibitem{MC90} Morozova E A and Chentsov N N 1990
{\sl Itogi Nauki Tehniki} {\bf 36} 135

\bibitem{Pe96} Petz D 1996 {\sl Linear Algebr. Appl} {\bf 244}, 81

\bibitem{BAR86} Balian R, Alhassid Y and Reinhardt H 1986
{\sl Phys. Rep} {\bf 131} 1

\bibitem{GS01} Grasselli M R and Streater R F 2001
{\sl  Inf. Dim. Analysis, Quantum Prob.} 
{\bf  4}, 173-182

\bibitem{KA80} Kubo F and Ando T 1980 
{\sl Math. Ann.} {\bf 246} 205

\bibitem{LR99} Lesniewski A and Ruskai M B 1999
{\sl J. Math. Phys.} {\bf 40} 5702

\bibitem{BS01}  Byrd M S and Slater P B 2001
Phys. Lett. {\bf A 283} 152

\bibitem{ZS01} {\.Z}yczkowski K and Sommers H-J  2001
 {\sl J. Phys.} {\bf A 34} 7111

\bibitem{Hu63}
Hua L K 1963 {\sl Harmonic Analysis of Functions of Several Variables
in the Classical Domains}  
(Providnece: American Ma\-the\-matical Society);
Chinese original 1958, Russian translation, Moskva 1959

\bibitem{Fu01} Fuiji K 2002 {\sl J. Appl. Math.} {\bf 2} 371

\bibitem{BST02} Boya L J,  Sudarshan E C G and Tilma T 2002
Volumes of compact manifolds
{\sl Preprint} math-ph/0210033 

\bibitem{TS02} Tilma T  and Sudarshan E C G 2002
{\sl J. Phys.} {\bf A 35} 10467

\bibitem{Me91}  Mehta M L 1991  {\sl Random Matrices}, II ed.
(New York: Academic)

\bibitem{Fy02} Fyodorov Y V 2002 
{\sl Nucl. Phys.} {\bf  B 621}  643

\bibitem{Di99b}  Dittmann J 1999 
 {\sl J.Geom.Phys.} {\bf 31} 16 

\bibitem{Sl99}  Slater P B 1999 
 {\sl J. Phys.} {\bf A 32} 5261

\bibitem{Sl02}  Slater P B 2002
{\sl Quantum Inf. Proc.} {\bf 1} 397

\bibitem{Sl03}  Slater P B 2003
{\sl Preprint}  quant-ph/0306132

\bibitem{EK95}  Eynard B and  Kristjansen C 1995
    {\sl Nucl. Phys.} {\bf  B455} 577

\bibitem{SZ04} Sommers H--J and {\.Z}yczkowski K  2004
{\sl J. Phys.} {\bf A 37} 8457

\bibitem{Zy99} {\.Z}yczkowski K 1999
{\sl Phys. Rev.} {\bf A 60} 3496
 

\end{thebibliography}
\end{document}